\documentclass[11pt,twoside]{article}
\usepackage{asp2014}

\aspSuppressVolSlug
\resetcounters

\begin{document}

\title{Cost Management on Commercial Cloud Platforms}

\author{G.~Bruce~Berriman$^1$, William O'Mullane$^2$, Arik Mitschang$^3$, and Ivelina
Momcheva $^4$ }
\affil{$^1$Caltech/IPAC-NExScI, Pasadena, CA 91125, USA; \email{gbb@ipac.caltech.edu}}
\affil{$^2$Rubin Observatory Project Management Office, Tucson, AZ  85721, USA}
\affil{$^3$Johns Hopkins University, Baltimore, MD 21218, USA}
\affil{$^4$ Space Telescope Science Institute, Baltimore, MD 21218, USA}


\paperauthor{G.~Bruce Berriman}{gbb@ipac.caltech.edu}{0000-0001-8388-534X}{Caltech}{IPAC/NExScI}{Pasadena}{CA}{91125}{USA}
\paperauthor{William O'Mullane}{womullan@lsst.org}{0000-0003-4141-6195}{ }{Rubin Observatory Project Management Office}{Tucson}{AZ}{85721}{USA}
\paperauthor{Arik Mitschang}{arik@jhu.edu@} {0000-0001-9239-012X} {Johns Hopkins University}{IDIES}{Baltimore}{MD}{21218}{USA}
\paperauthor{Gerard Lemson}{glemson1@jhu.edu}{0000-0001-5041-2458}{Johns Hopkins University}{IDIES}{Baltimore}{MD}{21218}{USA}
\paperauthor{Ivelina Momcheva}{imomcheva@stsci.edu}{0000-0003-1665-2073}{Space Telescope Science Institute}{ }{Baltimore}{MD}{21218}{USA}




  
\begin{abstract}

Commercial cloud platforms are a powerful technology for astronomical research. Despite the benefits of cloud computing---such as on-demand scalability and reduction of systems management overhead---confusion over how to manage costs remains, for many, one of the biggest barriers to entry. This confusion is exacerbated by the rapid growth in services offered by commercial providers,  by the growth in the number of these providers, and by storage, compute, and I/O  metered at separate rates---all of which can change without notice. As a rule, processing is very cheap, storage is more expensive, and downloading is very expensive. Thus, an application that produces large image data sets for download will be far more expensive than an application that performs extensive processing on a small data set. This Birds of a Feather (BoF) session aimed to quantify the above statement by presenting case studies of the costing of astronomy applications on commercial clouds that covered a range of processing scenarios; these presentations were the basis for discussion by the attendees.

\end{abstract}

\section{Introduction: Why Is Cost Management on the Cloud Difficult?}
 Cost management of cloud services can be complex and difficult because pricing is variable, with separate rates for compute, storage, data movement and use of APIs; because pricing changes without notice; because an abundance of providers now offer an exceptionally diverse range of services; and because billing practices can appear opaque. The authors of this paper have, however, acquired considerable collective expertise in managing costs. This paper aims to pass on this expertise for the benefit of the broader community, and represents the findings of a Birds of a Feather (BoF) session held as part of the ADASS XXX virtual conference.

The BoF presented five case studies of the costing of astronomy projects on commercial clouds:

\begin{itemize}
\item  W. O'Mullane "Hosting the Rubin Observatory Interim Data Facility on a cloud platform" (hereafter, WOM)

\item A. Mitschang "SciServer on a cloud platform" (hereafter, AM)

\item I. Momcheva "Managing cloud services at STScI" (hereafter, IMO)

\item B. Berriman "Creating TESS Mosaics on AWS" (hereafter, BBT)

\item B. Berriman "Cost Management Workshop at IPAC" (hereafter, BBC)

\end{itemize}
These presentations were followed by a discussion of cost management among the 100 persons attending. What follows summarizes the conclusions of the presentations and of the discussion.  All prices quoted below refer to prices in effect at the dates of provisioning. There is no guarantee that such prices remain in effect; nor does any pricing quoted constitute an endorsement of a provider.

\section{How Much Will It All Cost: Getting Started}

So far, at least in the US, funding agencies for astronomy and space science have not completed negotiations with cloud providers to offer special pricing for their grantees. At present, individual institutions or projects must negotiate themselves (WOM, IMO). They may also be able to take advantage of third-party "holding organizations" such as Cloud Bank. This pilot project manages cloud accounts for grantees for some NSF awards to avoid the high overheads imposed by many US universities. Such overheads are generally not imposed on purchases of on-premises hardware and can make total cloud costs prohibitively high.

It is important to understand what compute, storage, backup, database, and API or toolkit resources an application requires during the period it will run on the cloud. While many resources such as computing are inexpensive, others resources such as Amazon Elastic File Storage and egress of very large volumes out of the cloud can prove expensive (BBT, BBC). So can moving data across regional cloud data centers if held in different accounts, but this can be mitigated by using a third-party to manage them. Services such as Amazon Virtual Private Cloud (VPC) endpoints connect to the cloud without an internet gateway and can considerably reduce data movement costs.

The type of application needs to be considered as well. Cloud platforms are built with commodity hardware and are most suitable for embarrassingly parallel applications, rather than HPC or HTC applications, which require high-speed network connections. Moreover, be aware that experimenting with data types not yet well supported can lead to considerable expense; the example of memory mapping n-dimensional files is a case in point discussed during the BoF.

\section{Enterprise Price Negotiations}
It is crucial to perform due diligence and estimate the costs for all aspects of a data center. BBC described the need to consider overhead costs, availability marginal cost, sizing and flexibility, data transfer (ingress and egress), networking and firewalls, monitoring, and systems administration. There is no one-size-fits-all solution. For the case of the IPAC data center, with over 1 PB, Amazon Web Services (AWS) proved superior for availability, for sizing and flexibility, and for networks and firewalls, but IPAC was cheaper for overheads, marginal cost, and data egress; monitoring and systems administration were a tie.

In spring 2020, Wikipedia reported there were 285 cloud platforms offering a wealth of services, each with their own pricing. Some, such as the "big three"---AWS, Microsoft Azure and Google Cloud---can be considered "full service providers," while others may specialize in niche services. If, however, all that is needed is off-premises storage, a dedicated cloud storage provider may be the best choice; Wikipedia listed 143 of these in spring 2020. BBC pointed out that, for example, Wasabi offers "S3"-type object storage at a rate much cheaper than AWS, without egress charges, but it does not offer compute services. BBC also pointed out the cost of back-up storage for IPAC's data on AWS Glacier was comparable to on-premises backup, but may prove cheaper in total because on-premises maintenance costs are eliminated.

The cost estimates, once complete, provide a basis for negotiations with providers. This is best done by a dedicated cost management team familiar with the technical needs as well as business and financial management, and by engaging directly with providers.

WOM and IMO presented two examples of the results of negotiations conducted by the Rubin Observatory and the Space Telescope Science Institute (STScI). WOM described how, after thorough prototyping experiments with Google Cloud and AWS, the Rubin Observatory set up an interim data facility with Google Cloud as the provider for three years. 

IMO described how STScI used AWS through a third-party provider until 2018, and then engaged in major negotiations to establish an enterprise agreement with AWS. The best practices that emerged in negotiating discounts included: an upfront payment discount (“savings plan”) of 50-60\% (best for predictable workloads and a plan for storage costs upfront with commitment for a 10-20\% discount. The option to add credits can be added to discounts and options to waive some egress costs---typically 15\% of annual spending (this is now provided as standard by Google to Internet2\footnote{https://internet2.edu/} members) can be negotiated.

\section {What of Lonely Astronomers?}

Astronomers without access to negotiated prices have opportunities to avoid paying retail rates. IMO and BBT described these practices in detail:

\begin{itemize}
    \item Do an on-premises pilot project if possible, to understand the data and resources needed (BBT). 
    \item Understand workflow needs and how they translate to cloud services. How long will you need cloud services? Do you need to reprocess periodically? The TESS example demonstrated the need for large storage volume of 150 TB over the life of the project. 
    \item Use free or low-cost tiers to do the set-up. 
    \item Use the AWS Command Line Interface (CLI) to investigate compute resource management before, for example, building compute clusters.
    \item Consider requesting research credits to defray part of the costs.
    \item Consider using public data sets programs, which provide public serving of data of wide interest (though not for private data).
    \item Use "spot" pricing where possible, which offers very cheap rates for unused capacity. Jobs may stop if demand rises; jobs are saved and can be restarted.
    \item If there is adequate demand, use reserved server instances. AM showed that for the SciServer science platform, this can reduce annual charges by over 45\% compared with on-demand instances.
    \ item Consider infrastructure-as-code tools (such as Terraform) to simplify creation and destruction of cloud resources.
    \item Use serverless mode, whereby users submit their code and the provider runs and optimizes the server. Costs only accrue when jobs are running.
    \item Use S3 object storage whenever possible. Allow enough time to overcome the latency in S3 storage. Explore different S3 tiers for data that are not accessed frequently.
    \item Learn how to use cost calculators offered by providers. With some practice, they yield estimates that are good to 80\%.
    \item Ask support for help provisioning the resources you need.
\end{itemize}

A number of attendees proposed holding cost management workshops at national astronomy conferences. A repository of useful cost management tools and documentation was also thought valuable; an organization such as Cloud Bank may ultimately provide this function.

A further option is to use national or private clouds, though they may not be as powerful or full featured as commercial clouds. Examples are European Open Science Cloud (Europe), Nectar (Australia), and TACC Jetstream (US). They are free to qualified users, at the expense of writing a proposal. 

\section{Monitoring Costs and Resource usage: Avoiding Fiscal Black Holes }

Diligent attention to several procedures can avoid unnecessary costs and prevent financial headaches. IMO described them:

\begin{itemize}
    \item Do monitor costs daily.
    \item Do clean up jobs and do not forget about them. 
    \item Do run a test/prototype first and examine the changes in costs before scaling up. 
    \item Do understand the long tail of costs; avoid having jobs run for a long time. 
    \item Do use standard instances before customizing; customize or scale up only when necessary.
    \item Do think carefully about the storage costs. 
    \item Do download only the data you need. If you are downloading everything, re-think your cloud approach. Downloading large volumes of data can be very expensive (BBT, BBC). 
    \item Do set up all the billing alarms.
    \item Do automate shut down of a workflow once a spending cap is reached.
\end{itemize}




\end{document}